\documentclass[reprint,prc,superscriptaddress,nofootinbib]{revtex4-2}
\usepackage{amsmath,amssymb,graphics,graphicx,bbm,multirow,booktabs,tabularx}

\usepackage{hyperref}
\usepackage{xcolor}
\hypersetup{colorlinks,breaklinks,
  linkcolor=blue,urlcolor=blue,
  anchorcolor=blue,citecolor=blue}

\usepackage{lineno}
\newcommand{\nubb}   {\ensuremath{0\nu\beta\beta}}
\newcommand{\mbb}    {\ensuremath{m_{\beta\beta}}}

\newcommand{\mmio}   {\ensuremath{(m_{\beta\beta}^{min})_{\text{IO}}}}
\newcommand{\Thl}    {\ensuremath{T_{1/2}}}

\newcommand{\Ge}     {$^{76}$Ge}
\newcommand{\Mo}     {$^{100}$Mo}
\newcommand{\Xe}     {$^{136}$Xe}

\begin{document}

\title{Testing the Inverted Neutrino Mass Ordering with Neutrinoless Double-Beta Decay}

\author{Matteo Agostini}
\email{matteo.agostini@ucl.ac.uk}
\affiliation{Department of Physics and Astronomy, University College London, Gower Street, London WC1E 6BT, UK}
%\affiliation{Physik-Department E15, Technische Universit\"at M\"unchen, 85748 Garching, Germany}

\author{Giovanni Benato}
\email{giovanni.benato@lngs.infn.it}
\affiliation{INFN, Laboratori Nazionali del Gran Sasso, 67100 Assergi, L'Aquila, Italy}

\author{Jason A. Detwiler}
\email{jasondet@uw.edu}
\affiliation{Center for Experimental Nuclear Physics and Astrophysics, and Department of Physics, University of Washington, Seattle, WA 98115 - USA}

\author{Javier Men\'{e}ndez}
\email{menendez@fqa.ub.edu}
\affiliation{Department  of  Quantum  Physics  and  Astrophysics  and  Institute  of Cosmos Sciences, University of Barcelona, 08028 Barcelona, Spain}

\author{Francesco Vissani}
\email{vissani@lngs.infn.it}
\affiliation{INFN, Laboratori Nazionali del Gran Sasso, 67100 Assergi, L'Aquila, Italy}
\affiliation{Gran Sasso Science Institute, 67100 L'Aquila, Italy}

\date{\today}

\begin{abstract}
  We quantify the extent to which future experiments will test the existence of neutrinoless double-beta decay mediated by light neutrinos with inverted-ordered masses.
  While it remains difficult to compare measurements performed with different isotopes, we find that future searches will fully test the inverted ordering scenario, as a global, multi-isotope endeavor.
  They will also test other possible mechanisms driving the decay, including a large uncharted region of the allowed parameter space assuming that neutrino masses follow the normal ordering. 
  \end{abstract}

\maketitle %\maketitle must follow title, authors, abstract

Neutrino oscillation~\cite{Kajita:2016cak,McDonald:2016ixn,Eguchi:2002dm,Zyla:2020zbs} proves that the neutrino has mass and the Standard Model of particle physics is incomplete.
The unknown origin of the neutrino mass has drawn enormous attention to neutrinoless double-beta (\nubb) decay,
a matter-creating nuclear transition in which two neutrons decay simultaneously into two protons,
emitting only two new electrons and no antineutrinos~\cite{Furry:1939qr}.
The discovery of \nubb\ decay would establish that the neutrino is its own antiparticle and has a Majorana mass~\cite{Schechter:1981bd}.
It would also mark the first observation of a lepton-creating process, proving that neither lepton number ($L$) nor baryon minus lepton number ($B-L$) are symmetries of the Standard Model, as predicted by leading theories explaining %the origin of the neutrino mass~\cite{Vergados:2012xy} and
the matter-antimatter asymmetry of our universe~\cite{Fukugita:1986hr}.
Indeed, searching for \nubb\ decay is the most sensitive experimental approach to test Majorana neutrino masses and their associated $L$ violation. It is also a unique probe of new physics at ultrahigh energy scales not accessible by current accelerators~\cite{Deppisch:2015qwa,Peng:2015haa}.

Different physics mechanisms can lead to \nubb\ decay~\cite{Vergados:2012xy}. However, the exchange of light Majorana neutrinos interacting via standard, weak left-handed currents plays a special role. It is the only mechanism allowed by all theories predicting \nubb\ decay, and typically dominates the rate of the process~\cite{deGouvea:2007qla,Mitra:2011qr}.
Assuming that the decay is mediated by light neutrinos, its half-life is~\cite{Engel:2016xgb}
\begin{equation}
  \Thl^{-1}= G ~g_A^4~M^2~ \frac{\mbb^2}{m_e^2},
  \label{eq:thl}
\end{equation}
where $G$ is the phase-space integral, $g_A\simeq1.276$~\cite{Markisch:2018ndu} is the axial-vector coupling, $M$ is the nuclear matrix element,
and \mbb\ is the effective Majorana mass, normalized for convenience by the electron mass $m_{e}$.
The Majorana mass captures the physics of the exchanged neutrinos and is a function of the neutrino oscillation parameters, the neutrino mass eigenvalues $m_i$, and the Majorana phases: $\mbb= \bigl|\sum_i U^2_{ei}~m_i\bigr|$,
where $U_{ei}$ are the elements of the full 6-parameter PMNS matrix~\cite{Bilenky:2014uka}.

Neutrino oscillation measurements constrain the range of allowed \mbb\ values~\cite{Vissani:1999tu} and prove that $|U_{e3}^2| \ll |U_{e2}^2| < |U_{e1}^2|$ and $m_2^2 - m_1^2 \ll |m_3^2 - m_2^2|$ \cite{Super-Kamiokande:2016yck, Super-Kamiokande:2017yvm, T2K:2019bcf, NOvA:2019cyt, MINOS:2020llm, IceCube:2017lak, DayaBay:2018yms, RENO:2018dro, DoubleChooz:2019qbj, KamLAND:2013rgu}.
This implies that the effective Majorana mass is strictly larger than zero if neutrino masses follow the inverted ordering, i.e.~$m_3 < m_1 \lesssim m_2$. 
In this case, the lowest \mbb\ value, minimized with respect to the unknown Majorana phases and $m_3$, is given by
\begin{equation}
  \mmio = \bigl|U_{e1}^2\bigr|m_1 - \bigl|U_{e2}^2\bigr|m_2 - \bigl|U_{e3}^2\bigr|m_3,
  \label{eq:mmio}
\end{equation}
with $m_3 = \bigl|U_{e3}^2\bigr|/ \left( \bigl|U_{e1}^2\bigr| / m_1 - \bigl|U_{e2}^2\bigr| / m_2 \right)\approx 3$\,meV.
Using the latest values and uncertainties from the Particle Data Group~\cite{Zyla:2020zbs}, we obtain 
\begin{equation}
  \mmio = 18.4\pm1.3\,\text{meV},
\end{equation}
%for $m_3\approx2.9$\,meV, with 
whose uncertainty is dominated by the uncertainty on the solar mixing angle $\theta_{12}$.
Using the latest NuFIT results~\cite{Esteban:2020cvm}, we obtain $\mmio=18.6\pm1.2$\,meV.
%For $m_3 = m_3^{min}$ the masses obey $m_3 \ll m_1 \approx m_2 \approx \sqrt{\Delta m^2_{\rm atm}}$,
%where $\Delta m^2_{\rm atm}$ is the atmospheric neutrino mass-squared splitting.
%In this regime, Eq.~\eqref{eq:mmio} gives approximately $\mmio \approx \cos 2 \theta_{12}  \sqrt{\Delta m^2_{\rm atm}}$,
%showing that the uncertainty is dominated by the error on the solar mixing angle $\theta_{12}$.
The lower bound on \mbb\ corresponds to an upper bound on \Thl\ at the scale of $10^{27}$-$10^{28}$ years, depending on the value of the parameters in equation~\eqref{eq:thl}.
%Experiments with sensitivity reaching \mmio\  will establish 
%whether \nubb\ decay is mediated by the exchange of light neutrinos with inverted-ordered masses. 

In the last decades, a vast experimental program has been mounted to develop experiments  
with sensitivity reaching \mmio, able to exhaustively test whether \nubb\ decay is mediated by the exchange of light neutrinos with inverted-ordered masses~\cite{NLDBD,Giuliani:2019uno}. 
Mature designs are now available for multiple detection techniques, and the physics community is discussing how to proceed. As part of this process, 
the Astroparticle Physics European Consortium (APPEC) is updating its \nubb\ decay roadmap~\cite{Giuliani:2019uno} and the United States' Department of Energy has started a ton-scale-experiment portfolio review.
Conceptual designs are available for three experiments~\cite{CUPID:2019imh,LEGEND:2021bnm,nEXO:2018ylp}, whose construction can start as soon as funding is available. 
These experiments use different \nubb-decaying isotopes and detection technologies,
and can perform independent and complementary measurements.

%These efforts led to experiments reaching sensitivities at the level of 100\,meV,
%setting the stage for the selection of the most promising ideas for further investment.
%As part of this process, the United States’ Department of Energy has recently started a ton-scale-experiment portfolio review,
%with coverage of the inverted ordering parameter space identified as an explicit goal~\cite{}.
%Multiple projects able to discover \nubb\ decay at the bottom of the inverted ordering are being proposed.
%Conceptual designs are available for three experiments~\cite{CUPIDInterestGroup:2019inu,Abgrall:2017syy,Kharusi:2018eqi}
%whose construction can start as soon as funding is available.
%Each of these experiments relies on different \nubb-decaying isotopes and detection technologies.
%Such a variety is a richness for the field, allowing for independent and complementary measurements of \nubb\ decay. 

As mentioned above, observing \nubb\ decay would unambiguously demonstrate matter creation
and prove the Majorana nature of neutrinos. 
However, the conversion between \Thl\ and \mbb\ in Eq.~\eqref{eq:thl}
is not trivial and requires inputs from nuclear theory.
While the phase-space integral $G$ has been calculated with negligible uncertainty~\cite{Kotila:2012zza},
obtaining reliable nuclear matrix elements $M$ is challenging,
as it requires computationally intensive many-body calculations
and the evaluation of several operators~\cite{Engel:2016xgb,Cirigliano:2017tvr}. 
Four primary %nuclear
many-body methods have been historically used in the field: % have been used to compute these matrix elements:
the nuclear shell model (NSM)~\cite{Menendez18,Horoi16b,Coraggio20},
the quasiparticle random-phase approximation (QRPA) method~\cite{Mustonen13,Hyvarinen15,Simkovic18,Fang18,Terasaki20},
energy-density functional (EDF) theory~\cite{Rodriguez10,Vaquero14,Song17}, and
the interacting boson model (IBM)~\cite{Barea15,Deppisch:2020ztt}.
For each of these methods, several calculations have been performed  under different assumptions and approximations. The most recent results are listed in
Table~\ref{theTab}. They can differ by up to a factor of three for a given isotope, and significant differences are present even within each method.
The spread of values gives a rough idea of the many-body uncertainties on $M$ (additional uncertainty contributions are discussed below).
For some methods, calculations are not available for all isotopes.
%In particular, no NSM calculations are available for \Mo, where also the number of QRPA calculations is smaller than for other isotopes.
%This asymmetry needs to be taken into account when comparing quantitatively experiments using different nuclei.

\begin{table}[t]
  \caption{Nuclear matrix elements $M$ for \nubb\ decay mediated by light neutrinos, calculated with the NSM, QRPA, EDF, and IBM methods.
    The ranges correspond to the minimum and maximum values obtained with the same many-body method. 
  }
  \label{theTab}
  \begin{tabularx}{\columnwidth}{p{.15\columnwidth}p{.1\columnwidth}p{.23\columnwidth}p{.23\columnwidth}p{.23\columnwidth}}
    \toprule
    & Ref. & \Ge\ & \Mo\  & \Xe\ \\ %[2ex]
    \colrule
    \multirow{4}{*}{NSM}  & \cite{Menendez18}       & 2.89, 3.07  & --             & 2.28, 2.45  \\
                          & \cite{Horoi16b}         & 3.37, 3.57  & --             & 1.63, 1.76  \\
                          & \cite{Coraggio20}       & 2.66           & --             & 2.39           \\
                          & All                     & 2.66\,-\,3.57  & --             & 1.63\,-\,2.45  \\
    \colrule
    \multirow{6}{*}{QRPA} & \cite{Mustonen13}       & 5.09           & --             & 1.55           \\
                          & \cite{Hyvarinen15}      & 5.26           & 3.90           & 2.91           \\
                          & \cite{Simkovic18}       & 4.85           & 5.87           & 2.72           \\
                          & \cite{Fang18}           & 3.12, 3.40  & --             & 1.11, 1.18  \\
                          & \cite{Terasaki20}       & --             & --             & 3.38           \\
                          & All                     & 3.12\,-\,5.26  & 3.90\,-\,5.87  & 1.11\,-\,3.38  \\
    \colrule
    \multirow{4}{*}{EDF}  & \cite{Rodriguez10}      & 4.60           & 5.08           & 4.20           \\
                          & \cite{Vaquero14}        & 5.55           & 6.59           & 4.77           \\
                          & \cite{Song17}           & 6.04           & 6.48           & 4.24           \\ 
                          & All                     & 4.60\,-\,6.04  & 5.08\,-\,6.59  & 4.20\,-\,4.77  \\
    \colrule
    \multirow{3}{*}{IBM}  & \cite{Barea15}\footnote{With the sign change in the tensor part indicated in Ref.~\cite{Deppisch:2020ztt}.}
                                                    & 5.14           & 3.84          & 3.25           \\
%                                                   & 4.68(5.14)     & 4.22(3.84)     & 3.05(3.25)     \\
                          & \cite{Deppisch:2020ztt} & 6.34           & 5.08           & 3.40           \\
                          & All                     & 5.14\,-\,6.34  & 3.84\,-\,5.08  & 3.25\,-\,3.40  \\
    \bottomrule
  \end{tabularx}
\end{table}

The reach of \nubb\ decay experiments is conventionally expressed in terms of discovery and exclusion sensitivities on \mbb.
The discovery sensitivity corresponds to the smallest \mbb\ value for which an experiment has 50\% probability of observing a signal at 99.7\% confidence level (CL). 
The exclusion sensitivity corresponds to the median 90\%-CL upper limit that an experiment will set on \mbb\ assuming \nubb\ decay is not observable. 
As stated earlier, fully testing the inverted ordering scenario requires sensitivity to \mmio, accounting for its uncertainty. For discovery mode, this condition is met when the discovery sensitivity reaches the central value of \mmio: the \mmio\ uncertainty is symmetric, so the probability of lower or upper fluctuations is the same, and the 50\% probability for an observation is preserved. However, for exclusion mode, the \mmio\ uncertainty reduces the CL by a variable amount that depends on the experimental parameters, mainly the background statistical uncertainty. Therefore the exclusion sensitivity on \mbb\ cannot be used to set an experiment-independent condition corresponding to fully covering the inverted ordering scenario. The discovery sensitivity is anyway the most appropriate metric for searches aiming to discover a process.
Thus, reaching a discovery sensitivity of 18.4\,meV is the right concrete goal for experiments aiming to explore the full inverted ordering parameter space.

%To fully test the inverted ordering scenario, an experiment needs to be sensitive to \nubb\ decay with a \Thl\ value corresponding to $\mbb=\mmio$.
%To define a precise goal for future experiments, we can hence focus on testing this least favorable hypothesis, which corresponds to the weakest possible signal. 
%We revise the concepts of discovery and exclusion sensitivity, commonly adopted in the field to characterize an experiment's reach,
%to account for the fact that \mmio\ is not perfectly known.
%The discovery sensitivity is typically defined as the smallest \mbb\ value for which an experiment has 50\% probability of observing a signal with a 99.7\% %confidence level (CL) significance. To first order, the discovery sensitivity is not affected by the uncertainties on \mmio, as long as they are symmetric. In that %case, the probability of a \mbb\ lower or upper fluctuation about the central value is the same, so it does not alter the 50\% probability of performing an %observation.
%Conversely, incorporating the uncertainties into the exclusion sensitivity is not trivial. The exclusion sensitivity is defined as the largest \mbb\ value that an %experiment can exclude at 90\% CL with 50\% probability, assuming there is no signal. Uncertainties in \mmio\ will now reduce the CL by a variable amount that %depends on the experimental parameters, mainly the background statistical uncertainty. 

\begin{figure*}[t]
  \includegraphics[width=0.49\textwidth]{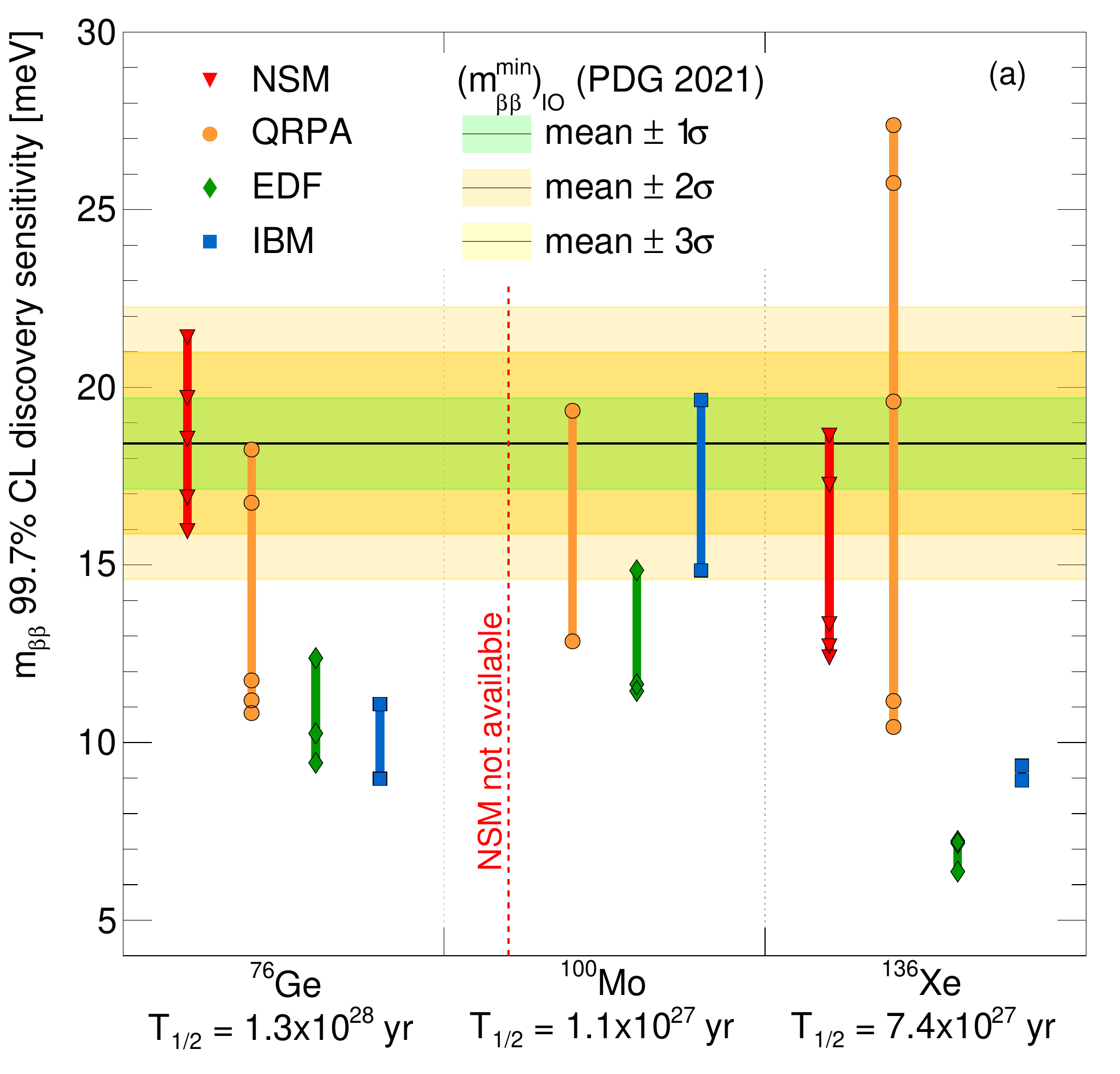}
  \includegraphics[width=0.49\textwidth]{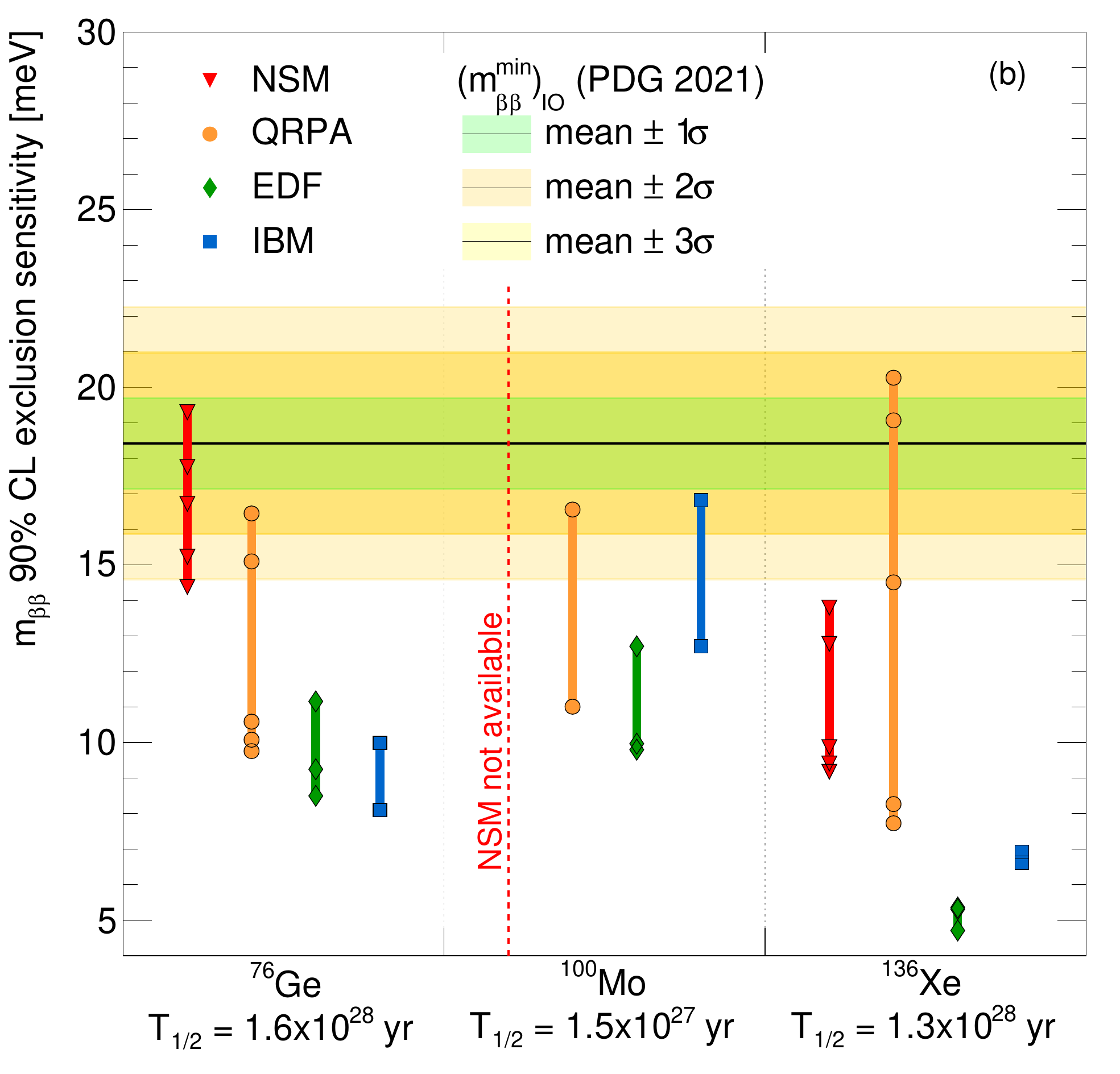}
  \caption{Comparison of \mbb\  99.7\%-CL discovery and 90\%-CL median exclusion sensitivities for different isotopes
    at stated half-life sensitivities~\cite{LEGEND:2021bnm,CUPID:2019imh,nEXO:2021ujk},
    grouped by nuclear many-body frameworks with matrix element ranges from Table~\ref{theTab}.
    The horizontal bands show the variation on \mmio\ under variation of the neutrino oscillation parameters.}
  \label{theFig}%
\end{figure*}
Figure~\ref{theFig} shows the discovery and exclusion sensitivities of proposed future experiments.
We converted the \Thl\ sensitivity values quoted by the LEGEND~\cite{LEGEND:2021bnm},
CUPID~\cite{CUPID:2019imh} and nEXO~\cite{nEXO:2021ujk} collaborations to \mbb\ values
using the nuclear matrix elements of Table~\ref{theTab}.
We group the calculations by many-body method to aid in the comparison of more consistent quantities.
Remarkably, these experiments all show sensitivity to measure a \nubb\ decay signal at the bottom of the inverted ordering parameter space.
Some many-body methods, such as EDF theory and IBM, give systematically larger $M$ values, pushing the \mbb\ sensitivity even lower.
QRPA calculations on the other hand give a broad range of results, partially due to the role of nuclear deformation in this framework.
The NSM provides $M$ values which are typically smaller than for the other methods,
but are not available for $^{100}$Mo. % due to the very demanding configuration space required for this method to describe nuclei in this region of the nuclear chart.

Comparing the performance of experiments using different isotopes is challenging
because of the large uncertainties affecting the nuclear matrix element predictions.
In particular, each many-body method uses different approximations, which are likely to result in a common over- or underestimation of the calculated $M$ values.
%Thus, we suggest that comparisons 
%consider one many-body method at a time, and are %avoided when one approach
%is not available for all isotopes. % (e.g NSM results for \Mo).
%
%While certainly more appropriate, 
Even comparisons considering one many-body method at a time can raise concerns.
The number of calculations available for each isotope and method can be significantly different, suggesting that not all many-body approaches are equally suitable for all isotopes. 
%The more calculations are available, the larger the spread of the \mbb\ values is likely to be. 
Consequently the range of $M$ values cannot be quantitatively interpreted as the uncertainty, which currently remains unknown. 
One might be tempted to make comparisons based on the central value of the \mbb\ sensitivity within a specific nuclear method, but this value can be disproportionately affected by a single outlier matrix element.
Weighted averages are also problematic, as the weight given to each calculation would be to some extent arbitrary.
Given the lack of objective criteria to compare experimental sensitivities in different isotopes, and the lack of a clear estimate of the uncertainties, we advocate to refrain from ranking experiments' reach quantitatively, and focus instead on the fact that we have a global, multi-isotope endeavor that will fully test the inverted ordering scenario.

%and the evaluation of several operators, including a recently identified contact term in heavy nuclei with still undetermined size and sign~\cite{Cirigliano:2017tvr,Cirigliano:2018hja,Cirigliano:2020dmx,Cirigliano:2021qko}.
%In addition, many-body calculations of operators dominated by the physics of spin -- as the leading contributions to $M$ -- typically require an ad-hoc ``quenching`` to match the experimental data, suggesting that some aspects of the many-body calculation or subleading operators have not been considered yet~\cite{Engel:2016xgb}.
%In this work we deal explicitly with the first of these three difficulties. We comment on the other two below.
%Figure~\ref{theFig} ignores uncertainties in nuclear matrix element calculations due to quenching and the contact term.

A broad effort to reduce uncertainties is ongoing within the nuclear theory community,
with significant advances made in the last few years. 
Ab initio calculations that incorporate wider nuclear correlations and two-body currents have recently succeeded
in predicting single $\beta$~decay rates~\cite{Gysbers:2019uyb} with no need for ``quenching'' -- an ad hoc reduction of the value of calculated matrix elements involving the nuclear spin required by less sophisticated calculations~\cite{Engel:2016xgb}. 
The first available ab initio \nubb\ matrix element calculations in medium-sized nuclei, supported by studies in lighter systems~\cite{Yao:2019rck,Novario:2020dmr},
indicate a relatively mild suppression by tens of percent with respect to the lower limit of the range given in Table~\ref{theTab}~\cite{Belley:2020ejd}.
Efforts are underway to improve the quality of these results, extend them to heavier nuclei, and include two-body currents at finite momentum transfers~\cite{Menendez:2011qq}. 
On the other hand, the contact term introduced in Refs.~\cite{Cirigliano:2017tvr,Cirigliano:2018hja},
which until recently went unrecognized, is a leading-order contribution to $M$.
Effective field theory and ab initio nuclear structure provide a scheme for estimating
this contribution~\cite{Cirigliano:2020dmx,Cirigliano:2021qko}. % which otherwise depends on a coupling which is not known experimentally.
A first study in $^{48}$Ca suggests that this term can enhance $M$ by about 40\% percent~\cite{Wirth:2021pij}, leading to a faster decay rate. %, but additional studies are needed to confirm this result.
In heavier systems, preliminary results suggest a roughly similar impact for all \nubb\ isotopes, only slightly dependent on the nuclear many-body method~\cite{Jokiniemi:2021qqv}. 
Complementary studies using e.g.~lattice QCD~\cite{Cirigliano:2020yhp,Davoudi:2020gxs} will test whether this claimed enhancement is robust.
If so, it may compensate the reduction in decay rate due to the inclusion of the ``quenching'' physics, leading to a picture similar to the one represented by Fig.~\ref{theFig}.
Thus, % even after resolving these theoretical difficulties, 
should the current theoretical results be confirmed,
the proposed global \nubb\ decay experimental effort % is still likely to 
would still fully probe the inverted ordering parameter space.

In this letter, we have focused on the inverted ordering scenario as a prominent goalpost for the proposed experimental \nubb\ decay program. 
%This milestones has been highlighted by the U.S. DOE/NSF Nuclear Science Advisory Committee’s Subcommittee on Neutrinoless Double Beta Decay~\cite{} and other funding bodies. 
However, the discovery power of these experiments is high even assuming other, equally reasonable scenarios.
By reaching a sensitivity of the order of tens of meV, these searches will probe a significant fraction
of the remaining parameter space for left-handed neutrino exchange even if neutrino masses follow the normal ordering.
Bayesian analyses suggest up to 50\% discovery probabilities for the normal ordering scenario~\cite{Agostini:2017jim,Caldwell:2017mqu},
and a non-vanishing discovery probability even assuming the most unfavorable value of the Majorana phases~\cite{Agostini:2020oiv}.
Significant advancement would also be made in probing the exchange of heavy mediators. %;
%some such models are amenable to complementary probes at accelerators, while for others 
For many such models, \nubb\ decay searches probe energy scales beyond the reach of current accelerator technology~\cite{Peng:2015haa}.
Additional physics mechanisms could completely change the parameter space of interest,
potentially even increasing the discovery power of future experiments~\cite{Cirigliano:2018yza, Tello:2010am, King:2013psa, Rodejohann:2011mu}.
In general, pushing \nubb\ decay sensitivity to increasingly large half-life values
explores uncharted territory, and new physics could manifest at any time.

\begin{acknowledgments}
  We would like to thank Christoph Wiesinger and Steven R.~Elliott for valuable discussions. 
  This work has been supported by the Science and Technology Facilities Council (Grant No.~ST/T004169/1),
  by the EU Horizon2020 research and innovation program under the Marie Sk\l{}odowska-Curie Grant Agreement No. 754496, by the Spanish MICINN through the ``Ram\'on y Cajal'' program with grant RYC-2017-22781,  the Italian Research Grant Number 2017W4HA7S ``NAT-NET: Neutrino and Astroparticle Theory Network'' under the program PRIN 2017 funded by MIUR, the AEI ``Unit of Excellence Mar\'ia de Maeztu 2020-2023'' award CEX2019-000918-M, the AEI grant FIS2017-87534-P
  and by the U.S.~DOE Office of Nuclear Physics under Grant Number DE-FG02-97ER41020.
\end{acknowledgments}

\end{document}